# Effects of Edge Oxidation on the Structural, Electronic, and Magnetic Properties of Zigzag Boron Nitride Nanoribbons

Dana Krepel and Oded Hod

Department of Chemical Physics, School of Chemistry, The Raymond and Beverly Sackler Faculty of Exact Sciences, Tel Aviv University, Tel Aviv 69978, Israel

Abstract

The effects of edge chemistry on the relative stability and electronic properties of zigzag boron nitride nanoribbons (ZBNNRs) are investigated. Among all functional groups considered fully hydroxylated ZBNNRs are found to be the most energetically stable. When an in-plane external electric field is applied perpendicular to the axis of both hydrogenated and hydroxylated ZBNNRs a spin-polarized half-metallic state is induced whose character is different than that predicted for ZGNRs. The onset field for achieving the half-metallic state is found to mainly depend on the width of the ribbon. Our results indicate that edge functionalization of ZBNNRs may open the way for the design of new nano-electronic and nano-spintronic devices.

The recent experimental realization of atomically thin, long, and narrow strips of graphene, often referred to as graphene nanoribbons (GNRs),[1-5] has triggered extensive experimental and theoretical investigations of their physical properties. Due to their unique structural, mechanical, electronic, and magnetic characteristics, GNRs have been identified as promising candidates for numerous potential applications including spintronic devices, gas sensors, and nano-composite materials.[6-9] The quasi-one-dimensional nature of GNRS results in the presence of reactive edges that may dominate their electronic and magnetic behavior.[10,11] Specifically, when grown along the zigzag axis the resulting zigzag graphene ribbons (ZGNRs) present pronounced electronic edge states[12-14] making them prone to covalent attachment of chemical groups that can significantly alter their electronic properties.[5,15-18] Within the simplest edge passivation scheme, hydrogen terminated ZGNRs were shown to exhibit a spin polarized semiconducting ground state with bandgaps that vary with the ribbon width.[14] This ground state is characterized by energetically degenerate α and β molecular spin orbitals that are spatially related via inversion symmetry. Here, opposite spin orientations localize at the two edges of the ZGNR and couple through the graphene backbone via an antiferromagnetic (AF) arrangement of spins on adjacent atomic sites. The application of an in-plane electric field, perpendicular to the ZGNR's axis, was shown to lift this bandgap degeneracy thus creating a perfect spin filter where electrons carrying one spin flavor present metallic behavior while their opposite spin counterparts exhibit wide bandgap semi-conductor characteristics.[19]

Hexagonal boron nitride (*h*-BN) is the inorganic analogue of graphite. The two materials are isoelectronic and their hexagonal lattices are isomorphic. In analogy to graphene, a single layer of *h*-BN has an equal number of $sp^2$ hybridized B and N atoms, which are covalently bonded in an alternating pattern. Nevertheless, due to the differences in electron affinities of the boron and nitrogen atoms, the B-N bonds have a considerable ionic character. This results in major differences both in their optimal interlayer stacking (graphite presents the Bernal ABA stacking whereas *h*-BN has the anti-eclipsed AA'A stacking)[20] and in their electronic properties (graphene is a semimetal whereas *h*-BN is an insulator)[21].

Because of these characteristics, *h*-BN has also attracted significant attention from the scientific community.[21-25] With this respect, the electronic and magnetic properties of boron-nitride nanoribbons (BNNR) with bare edges have been thoroughly investigated providing theoretical evidence that they are magnetic semiconductors with bandgaps that decrease with increasing ribbon width.[26,27] Half-metallicity has also been predicted to occur in zigzag BNNRs (ZBNNRs) with fluorine edge-decoration.[28] Furthermore, it was shown that terminating both the B and the N zigzag edges with oxygen and sulfur atoms gives rise to metallic behavior in these systems.[29]

Here, we present a first-principle computational study of the effects of edge oxidation on the relative stability, electronic properties, and half-metallic nature of ZBNNRs. To this end, we consider six different oxidation schemes of a 1.37 nm wide ZBNNR as shown in Fig. 1. We notate these oxidation schemes in accordance to their GNR counterparts namely:[5] partial hydroxylation (panel b, OH(I)), full hydroxylation (panel c, OH(II)), etheration schemes (panel d, Et(I), and e, Et(II)), partial ketonation (panel f, Kt(I)), and full ketonation (panel g, Kt(II)).

Our calculations have been carried out utilizing the GAUSSIAN suite of programs.[30] Spin-polarized calculations have been performed within the local density approximation (LDA), the generalized gradient approximation of Perdew et al. (PBE)[31,32], and the screened-exchange hybrid functional developed by Heyd, Scuseria, and Ernzerhof (HSE)[33-36] of density functional theory (DFT). We note that the latter functional approximation has been shown to reproduce experimental optical bandgaps of bulk semi-conductors and to describe the physical properties of wide variety of materials, including carbon nanotubes and graphene and its derivatives, with much success.[5,37-41] Unless otherwise stated the double-zeta polarized 6-31G$^{**}$ Gaussian basis set was used.[42]

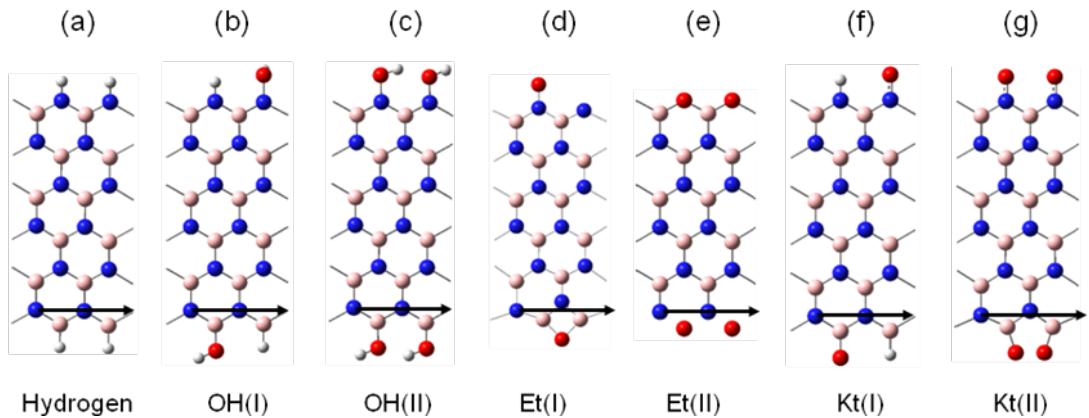

Figure 1: Optimized unit-cell geometries of different edge oxidation schemes studied in this work. The presented structures were obtained at the HSE/6-31G** level of theory. Color code: red, oxygen atoms; blue, nitrogen atoms; pink, boron atoms; grey, hydrogen atoms. Arrows represent the periodic direction of the ZBNNRs.

We start by discussing the optimized structures of the various systems considered. In Fig. 1 we present the relaxed geometries obtained at the HSE/6-31G** level of theory. For the hydrogen terminated system (Fig. 1(a)) we find that the N-H bonds (1.01 Å) are somewhat shorter than the B-H bonds (1.12 Å). Upon partial hydroxylation of both edges (Fig. 1(b)) the hydroxyl group attached to the boron edge remains in the plane of the ribbon whereas the hydroxyl group bonded to the nitrogen edge rotates to become perpendicular to the basal plane of the system. For the fully hydroxylated system the formation of a network of hydrogen bonds between adjacent OH groups causes the substituents to remain in the plane of the ribbon similar to the case of fully hydroxylated GNRs.[5,43] When both edges are partially decorated with ether groups (Fig 1(d)), the optimization procedure results in a structure where the boron edge exhibits an ether-like configuration whereas the nitrogen edge obtains a partial ketone-like structure. In Fig. 1(e) both edges are fully functionalized with ether-like groups where oxygen atoms replace the edge boron and nitrogen atoms. Here, the geometry optimization maintains the general structure while elongating the N-O bonds (1.49 Å) as compared to the B-O (1.39 Å) counterparts. A somewhat different picture arises upon partial ketonation (Fig. 1(f)) where the obtained N-O bonds (1.26 Å) are shorter than the B-O bonds (1.40 Å). Finally, when covering both edges with ketone-like groups, the optimized structure presents a ketonated nitrogen edge and a peroxide-like structure on the boron edge. For simplicity, we name the different structures according to the original functionalization scheme prior to geometry optimization.

Next, we study the relative stability of the different oxidized ribbons. As these structures have different chemical compositions, the cohesive energy per atom does not provide a suitable measure for the comparison of their relative stability. Therefore, we adopt the approach used in Refs. [5,37,44] where one defines a Gibbs free energy of formation $\delta G$ for a BNNR as follows:

$$\delta G(\chi_H, \chi_O) = E(\chi_H, \chi_O) - \chi_H \mu_H - \chi_O \mu_O - \chi_{BN} \mu_{BN} \qquad (1)$$

Where $E(\chi_H, \chi_O)$ is the cohesive energy per atom of a BNNR with given composition and dimensions, $\chi_i$ is the molar fraction of atom $i$ ($i$=H, O, or BN couple) in the ribbon satisfying the relation $\sum_i \chi_i = 1$, and $\mu_i$ is the chemical potential of the constituent at a given state. We choose $\mu_H$ as the binding energy per atom of the singlet ground state of the hydrogen molecule, $\mu_O$ as the binding energy per atom of the triplet ground state of the oxygen molecule, and $\mu_{BN}$ as the cohesive energy per BN couple of a single two-dimensional BN sheet, all calculated at the same level of theory as $E(\chi_H, \chi_O)$. This definition allows for an energy comparison between oxidized nanoribbons with different compositions, where negative values represent stable structures with respect to the constituents. For all oxidation schemes considered, we compared the closed-shell singlet spin state with the triplet spin state in order to identify the lowest energy spin configuration of each system.

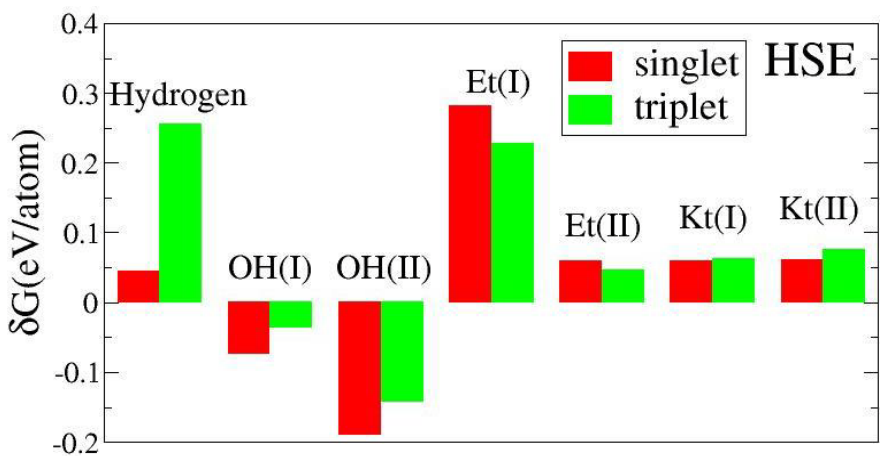

Figure 2: Ground state relative stabilities of the different oxidized systems studied (see Fig. 1) obtained via Eq. (1) at the HSE/6-31G** level of theory. Negative values indicate stable structures with respect to the constituents. Results of the singlet and triplet spin state calculations are presented in red and green bars, respectively.

In Fig. 2 we present the relative stabilities of the different oxidized ZBNNRs studied using the HSE functional approximation (LDA and PBE results are presented for comparison in Fig. S1 of the supporting information). As can be seen, the fully hydrogenated ribbon as well most of the oxidized ribbons considered, are found to be less stable than their corresponding constituents. Conversely, both hydroxylation schemes lead to considerable energetic stabilization of the ribbon's structure. We find that the most stable structure corresponds to the full edge-hydroxylation scheme (Fig. 1 (c)). As mentioned above, this enhanced stability is attributed to the hydrogen bonds formed between adjacent edge hydroxyl groups and is consistent with previous calculations on edge oxidized GNRs.[5] Furthermore, we note that for the fully hydrogenated system, as well as for both hydroxylated structures, the closed-shell singlet spin state is more energetically stable than the corresponding triplet state.

Having identified the most stable edge oxidation scheme, we now turn to study the electronic properties of the oxidized ZBNNRs considered. In Fig. 3 we present the bandgaps, calculated as energy differences between the lowest unoccupied (LUCO) and highest occupied (HOCO) crystalline Kohn-Sham orbitals, using the HSE functional approximation (LDA and PBE results are presented for comparison in Fig. S2 of the supporting information). Our calculations show that the bandgap obtained for the fully hydrogenated system is 5.72eV using the HSE functional and 4.24 eV using the PBE approximation, which is in good agreement with previous studies of similar structures (5.56eV [26] and 4.26eV [29], respectively). Interestingly, the two hydroxylation schemes present relatively small (~0.2 eV) modifications of the bandgap preserving the system's insulating character while maintaining the indirect nature of the bandgap. Furthermore, the lower hydroxyl density is found to slightly increase the bandgap whereas the higher hydroxyl density slightly decreases the bandgap with respect to the fully hydrogenated ZBNNR.

Markedly, the bandgap response towards the other edge oxidation schemes is highly diverse. Here, the triplet ground states of the Et(I) and Et(II) oxidation schemes lift the spin degeneracy such that for Et(I) the bandgap of one spin flavor is reduced by ~11% (0.62 eV) while the bandgap of the opposite spin electrons drops by 59% (3.36 eV). For the Et(II) oxidation scheme both up and down spin bandgaps reduce by ~62% thus turning the system into a wide bandgap semiconductor. For the Kt (I) and Kt (II) structures the HSE bandgaps further decrease turning the system metallic for the Kt (I) structure and a narrow bandgap semiconductor for the Kt (II) system. Notably, all these oxidation schemes, most of which are comparable in stability to the edge hydrogenated system, result in direct bandgap materials thus demonstrating that edge chemistry can be used as an efficient control scheme for tailoring the electronic properties of ZBNNRs.

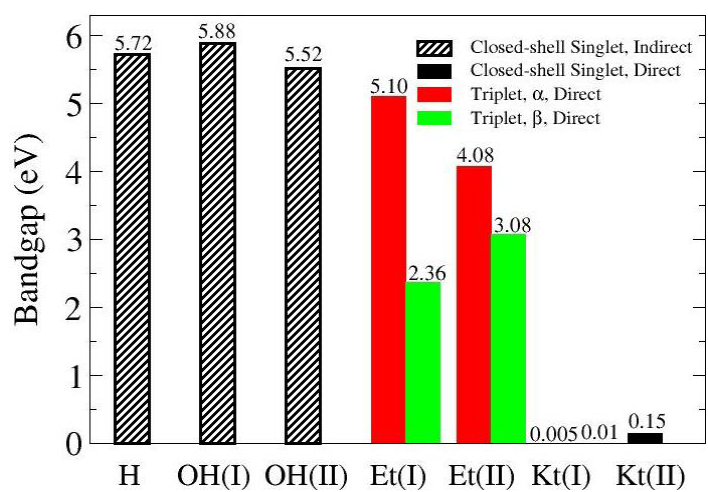

Figure 3: Bandap response to different oxidation schemes studied at the HSE/6-31G** level of theory. The singlet spin state is represented by black bars whereas the triplet spin state is represented by red (α spin electrons) and green (β spin electrons) bars. Full and striped bars represent direct and indirect bandgaps, respectively.

In order to gain better understanding of the varied influence of the different oxidation schemes on the electronic properties of the ZBNNR, we performed band structure and density of states (DOS) analysis.[45] In Fig. 4 we present the band structures along with the full DOSs (FDOS) and the partial DOSs (pDOS) of the B and N edges for the hydrogenated, hydroxylated, and a representative ketonated ZBNNRs. The edge PDOSs include only the contributions of the oxidation functional groups covalently bonded to the relevant edge atoms.

When examining the electronic properties of the hydrogen terminated system (Fig. 4(a)), it is found that the highest occupied crystalline orbital (HOCO) and lowest unoccupied crystalline orbital (LUCO) are associated with the central section of the ZBNNR. The N-edge contribution appears ~2 eV above the conduction band minimum (CBM) whereas the B-edge DOS appears ~2.5 eV below the valence band maximum (VBM). Partial hydroxylation (Fig. 4(b)) results in relatively minor changes of the HOCO and LUCO dispersion relations. From the PDOS analysis it is found that the N-edge now contributes some DOS to the low lying valence bands while the B-edge develops DOS at the high lying conduction bands. When the system is fully hydroxilated (Fig. 4(c)), qualitative changes in the HOCO and LUCO dispersion relations, including the location of the VBM and CBM, are evident. Here, the N-edge is found to contribute DOS at the HOCO and near the LUCO and the B-edge gains some DOS above the LUCO band. It should be noted that a similar picture appears for the α spin state of the less stable Et(I), Et(II) and Kt (II) oxidation schemes (see supporting information Fig. S3), whereas the β spin state of both schemes present relatively flat bands in the mid gap area. DOS analyses reveal these bands are contributed from the N-edge.

The partial ketonation scheme is found to p-dope the system by shifting the Fermi energy into the valence band (Fig. 4(d)). The screened hybrid HSE functional approximation predicts a small (~0.01eV) direct bandgap whereas both the PBE and LDA functionals predicts metallic behavior. Here, both ketonated B- and N-edges present pronounced DOS at the Fermi energy with clearly evident qualitative changes of the low energy band structure.

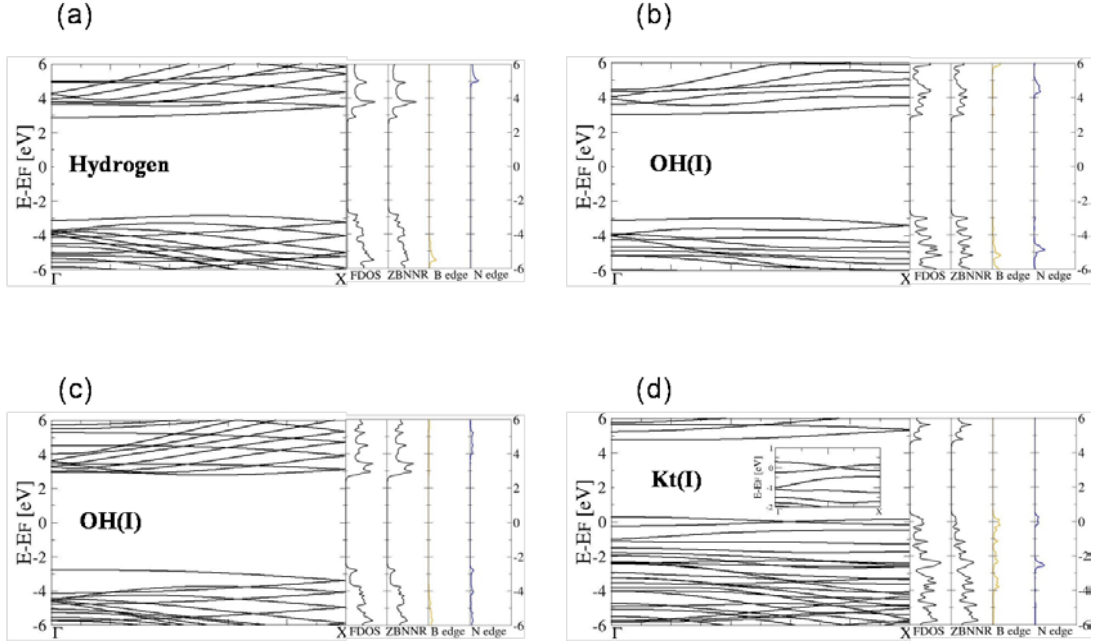

Figure 4: Band-structures, FDOS and PDOS of the (a) hydrogen terminated; (b) partially hydroxylated (OH(I) scheme); (c) fully hydroxylated (OH(II) scheme); and (d) partial ketonated (Ket(I) scheme) ZBNNR as calculated at the HSE/6-31G** level of theory. Fermi energies of all diagrams are set to zero. Inset of panel (d): zoom-in on the band-structure in the range of -2.0-1.0eV around the Fermi energy.

Similar to the case of GNRs, the effects of edge chemistry on the electronic properties of ZBNNRs is expected to alter their response towards the application of external fields. Previous studies have shown that bare ZBNNRs may present a rich spectrum of electronic characteristics ranging from metallic through half-metallic to semiconducting under the influence of external electric fields.[26] Thus, it would be interesting to investigate whether edge oxidation may be used to further control the response of ZBNNRs to such external perturbations. In what follows, we focus the discussion on the most stable (hydroxylated) decoration scheme while using the hydrogenated system as a reference.

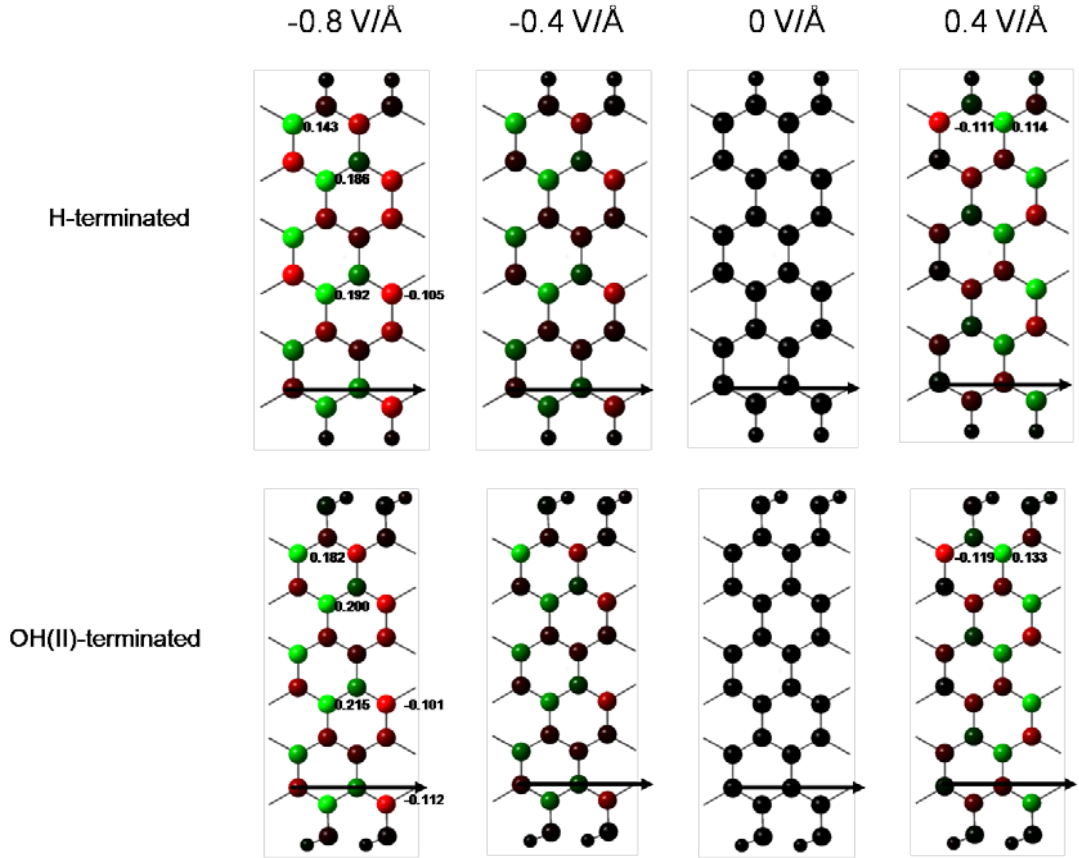

Figure 5: Ground state Mulliken atomic spin polarization calculated at the HSE/6-31G** level of theory for the hydrogenated (upper panels) and fully hydroxylated (lower panels) ZBNNRs at various external electric field intensities. Color code: red, α spin excess; green, β spin excess. Mulliken spin range is ±0.1 in all panels. Representative Mulliken spin values are given for completeness. Arrows represent the periodic direction of the ZBNNRs.

To this end, we present in Fig. 5 the ground state Mulliken atomic spin polarizations of the hydrogenated and the fully hydroxylated ZBNNRs obtained at the HSE/6-31G** level of theory at various in-plane electric field intensities applied perpendicular to the main axis of the ribbons. Since ZBNNR lack mirror symmetry with respect to their central axis not only the intensity but also the direction of the field has influence on the electronic properties of the system. Therefore, we consider both positive (pointing from the B- to the N-edge) and negative (pointing from the N- to the B-edge) electric fields. Opposite to the case of ZGNRs, where the ground state magnetization decreases with increasing external field intensity,[5] here, in the absence of an external field the ZBNNRs are found to be non-magnetic and the application of the external electric field induces spin polarization, regardless of its direction, both in the hydrogenated and the fully hydroxilated systems. As can be seen in Fig. 6, a quadratic-like decrease in the energetic stability of the closed-shell singlet spin state with respect to its open-shell counterpart is obtained with increasing field intensities. Remarkably, energy differences exceeding 1.1 eV/unit-cell were obtained for the strongest external field considered of +0.8 V/Å for the hydrogen terminated ZBNNR. The partially and fully hydroxylated systems presented energy differences of 1.0 and 0.7 eV/unit-cell, respectively, at this field intensity. For a positive field intensity of 0.4 V/Å we find that the replacement of edge hydrogen atoms with hydroxyl groups somewhat reduces the relative energetic stability of the open-shell singlet spin state.

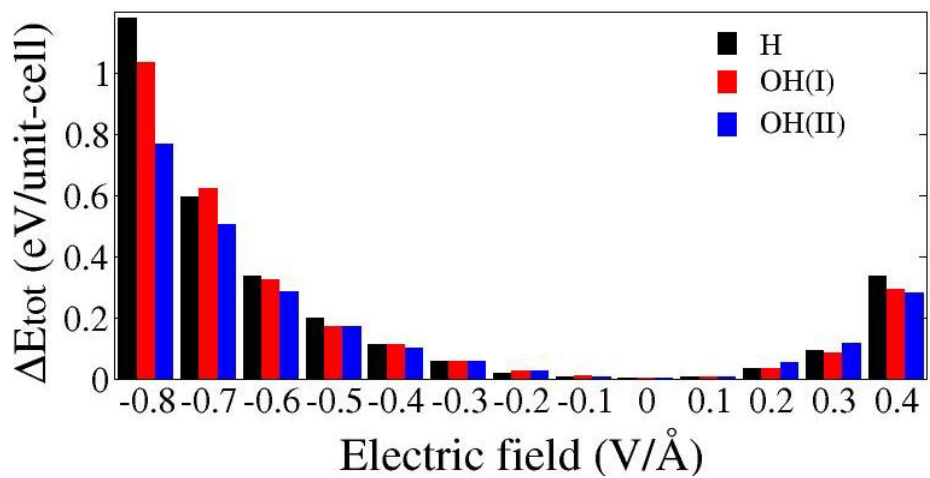

Figure 6: Energy differences between closed-shell and open-shell singlet spin state as function of external electric field intensity for the H, OH(I) and OH(II) edge decoration schemes obtained at the HSE/6-31G** level of theory.

With the understanding that an external electric field can drastically change the magnetic character of ZBNNRs we now turn to discuss its influence on their electronic properties. In Fig. 7 the band gap of hydrogenated and hydroxylated ZBNNRs as a function of the external field intensity is presented. As can be seen, at relatively low intensities the bandgap associated with both α and β spin electrons are insensitive to the application of the external electric field. At a certain critical field strength, that varies with the edge passivation scheme, the α bandgap, rapidly drops to zero while the β bandgap remains intact resulting in a half-metallic state. Interestingly, unlike the case of ZGNRs, the half-metallicity onset field in ZBNNRs is weakly dependent on the edge hydroxylation scheme. Nevertheless, a careful examination reveals that the half-metallicity onset field is different for the positive and negative field directions. This may be rationalized by the fact that the B- and N-edges of the ZBNNR are oppositely charged and therefore an inherent positive electric field is induced in the system. Hence, at the positive direction a lower external field intensity is required to induce half-metallicity.

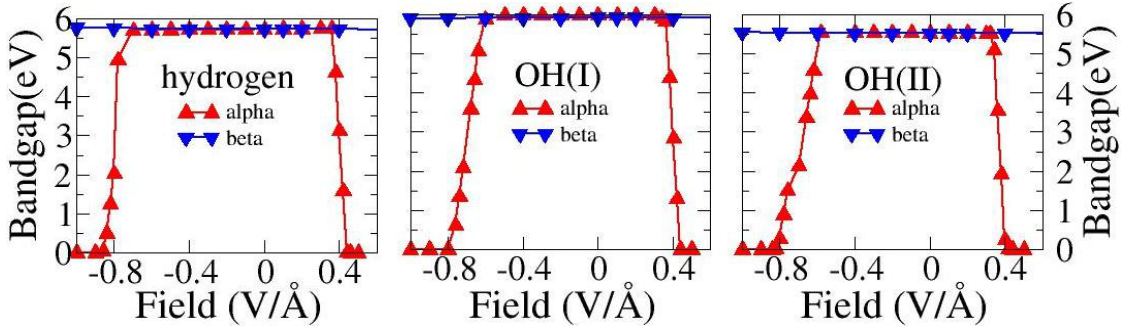

Figure 7: Spin polarized band gap as a function of the external electric field intensity of the hydrogenated (left panel), and partially (middle panel) and fully (right panel) edge-hydroxylated ZBNNRs obtained at the HSE/6-31G** level of theory.

In order to better understand the origin of the field induced half-metallic state observed in the ZBNNRs studied we plot in Fig. 8 the $\alpha$ electrons band structure along with the full $\alpha$ DOS and the B- and N-edges pDOSs of the hydrogen terminated ZBNNR at various electric field intensities. We note that the band structure associated with the $\beta$ electrons is virtually insensitive to the external electric field in the intensity regime studied. As discussed above, in the absence of an external electric field the ground state of the system is of closed-shell insulating character with a wide HSE bandgap of 5.72 eV. At a field intensity of -0.8 V/Å two energy bands originating from the field-less conduction band penetrate the bandgap region with a parabolic-like dispersion relation thus considerably reducing the value of the bandgap. The partial DOS analysis reveals that these bands are associated with the central section of the ribbons and not with its edges consistent with the Mulliken spin polarization plots presented in Fig. 5. As the field intensity is further increased to -0.9 V/Å one of the two parabolic energy bands crosses the Fermi energy and the system becomes half-metallic. The main influence of the external field on the edge DOS is an up-shift with respect to the Fermi energy. A similar qualitative picture with some quantitative differences is obtained when reversing the direction of the externally applied electric field. We note that similar results are obtained for the fully hydroxilated systems (see supporting information Fig. S4). This behavior is completely different from that observed in ZGNRs where a field induced half-metallic state is achieved due to opposite local gating of the edge α and β spin DOSs.[14]

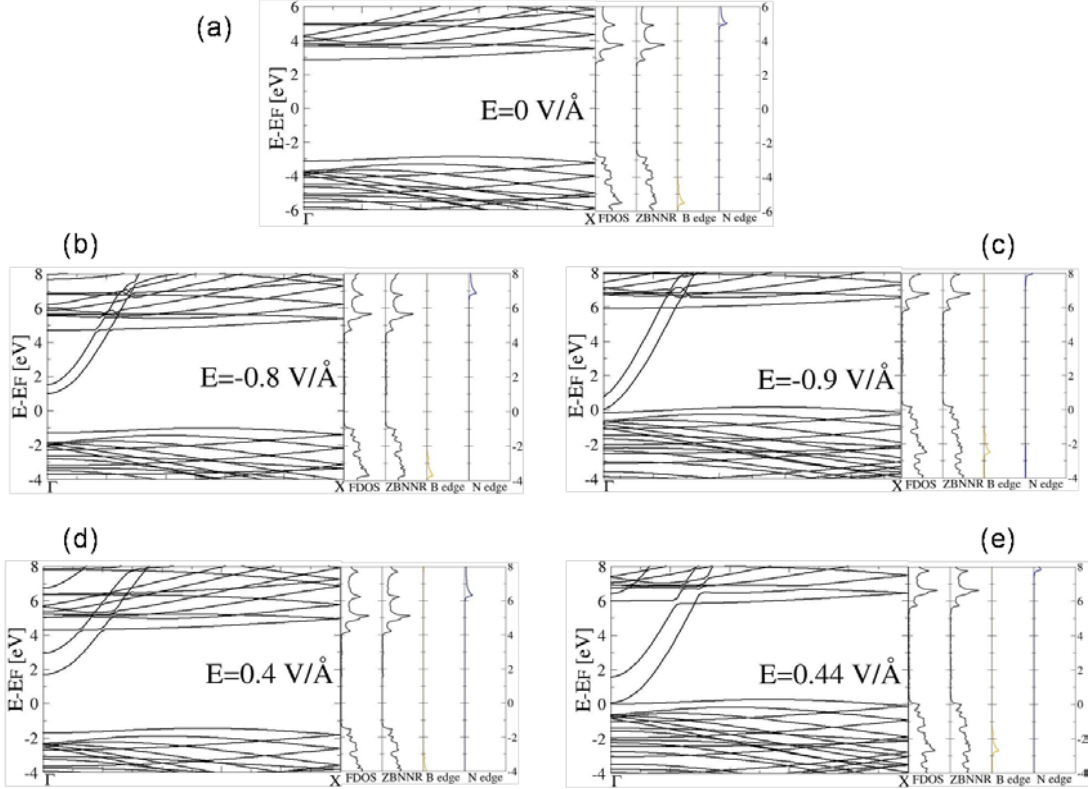

Figure 8: Band-structures, FDOS and PDOS of the α spin state of the H-terminated ZBNNR at the following fields: (a) 0V/Å, (b) -0.8 V/Å (c) -0.9V/Å (d) +0.4V/Å, and (e) +0.44 V/Å as calculated at the HSE/6-31G** level of theory. Fermi energies of all diagrams are set to zero.

Finally, to verify the general nature of our results we have performed similar calculations on hydrogenated and hydroxylated ZBNNRs of different widths. Previous studies of different spin states of bare BNNRs suggested that the energy difference between different spin states should vanish above a certain ribbon width.[26] Furthermore, the onset field for achieving a half-metallic state was shown to somewhat decrease with increasing bare ribbon width.

Here, we perform calculations on three edge decorated unit-cells of consecutive widths. We annotate these unit-cells by (*N*x*M*) where *N* stands for the number of zigzag chains along the width of the ribbon and *M* for the number of boron-nitride pair chains along its zigzag edge. Using this notation we study the (4×4) and (8x4) unit-cells and compare the result to those presented above for the (6x4) unit cell (see Fig. 9), thus creating ribbons of the following widths: 0.94nm, 1.37nm and 1.81nm.

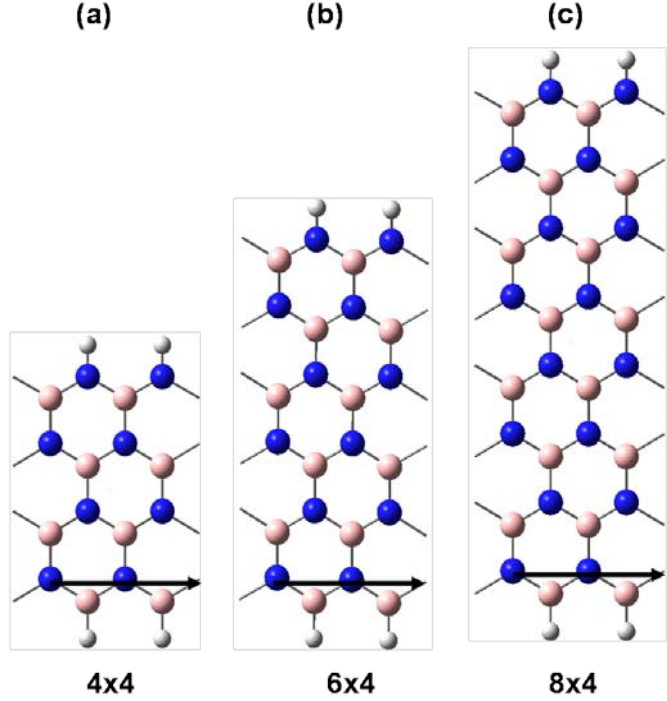

Figure 9: Schematic diagrams of the hydrogenated (a) (4×4), (b) (6×4) and (c) (8×4) zBNNRs unit-cells. Arrows represent the periodic direction of the ZBNNRs.

To study the influence of the width of the ribbons on their relative energetic stability we repeat the analysis performed above according to eq. (1) for the edge-hydrogenated and partially and fully hydroxylated ribbons of various widths. The results for the different unit-cell dimensions are summarized if Fig. 10. We find that increasing the width of the ribbon results in an increase in its stability while generally maintaining the relative stability ordering of the three edge-decoration schemes.

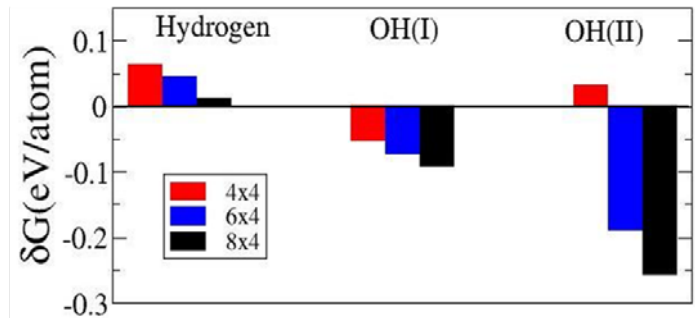

Figure 10: Relative stabilities of the H, OH(I) and OH(II) oxidation schemes (see Fig. 1) as a function of ZBNNR width obtained via eq. (1) at the HSE/6-31G** level of theory and the closed-shell singlet spin state. Negative values indicate stable structures with respect to the constituents.

The width of the ribbon was also found to influence its response towards the application of an external electric field. In Fig. 11 we compare the bandgap dependence on the field intensity for the three ribbon widths considered with hydrogen and hydroxyl edge decoration. As can be seen, all systems considered present a similar zero-field bandgap of 5.5-6 eV. Nevertheless, upon the application of a positive external field the onset of half-metallicity shifts towards higher field intensities as the width of the ribbon is increased. An opposite picture arises for the negative field direction where the onset of half-metallicity shifts towards lower field intensities with increasing ribbons width. A similar behavior was also found in both ZGNRs and bare ZBNNRs.[14,26] This width dependence may be explained using the simple model discussed above according to which the charge polarization between the B- and N- edges forms an intrinsic field that enhances or reduces the effect of the external field depending on its direction. As the width of the ribbon increases, the onset field intensity to obtain a half-metallic state in the negative (positive) direction is decreased (increased). This indicated that the intrinsic effective field decreases with increasing ribbon width. Nonetheless, we find that the qualitative nature of the response is similar for all systems studied thus further supporting the general nature of our predictions.

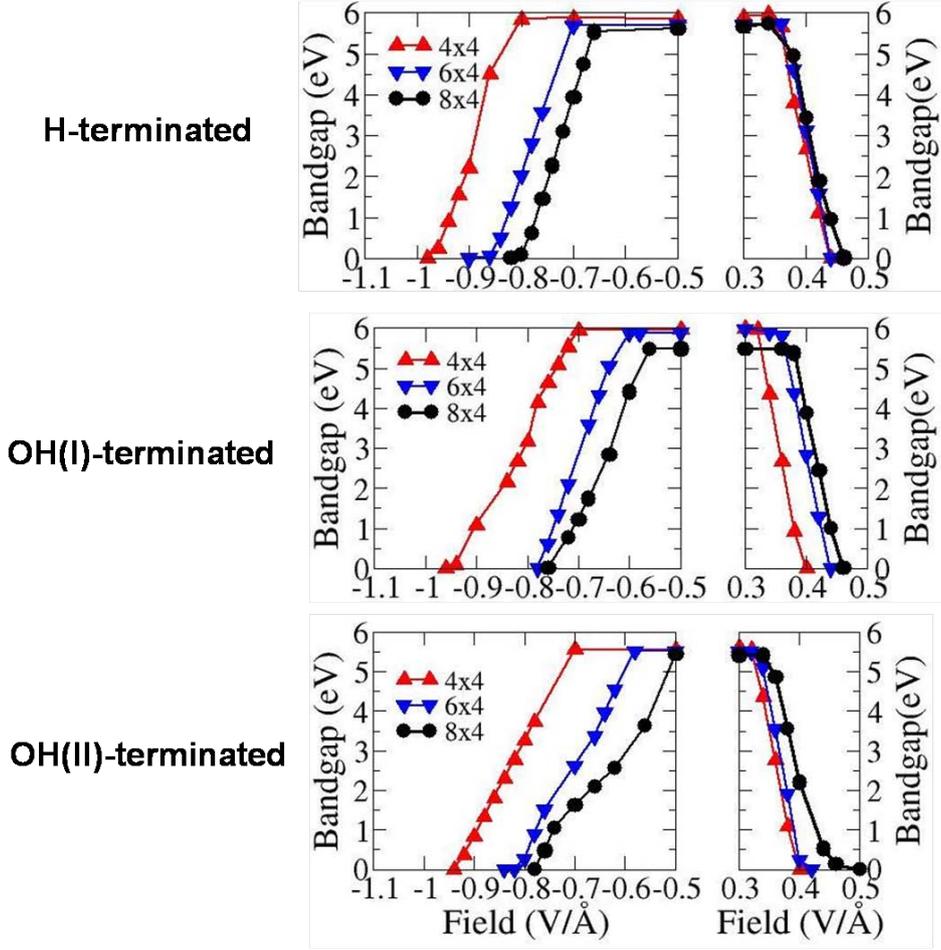

Figure 11: α spin bandgaps as a function of the field intensity and direction of the hydrogenated (upper panels) and partially (middle panels) and fully (lower panels) hydroxylated ZBNNRs of varying width obtained at the HSE/6-31G** level of theory.

In summary, we have studied the relative stability, electronic properties, and response to external electric field perturbations of ZBNNRs with various edge-oxidation schemes. It was found that chemical functionalization of the edges may considerably influence the relative stability of the system. Among all functional groups considered the fully hydroxylated ZBNNR was found to be the most energetically stable. This stability was found to enhance with increasing ribbon width. The influence of edge-functionalization on the ground state electronic structure was found to strongly depend on the chemical nature of the decorating group, where electronic character ranging from insulating to metallic was obtained for different oxidation schemes. The application of an external in-plane electric field perpendicular to the axis of edge-hydrogenated and hydroxylated ZBNNRs was found to induce a spin-polarized half-metallic state, whose nature is completely different than that observed in ZGNRs. The onset field for achieving the half-metallic state was found to mainly depend on the width of the ribbon with moderate sensitivity towards edge hydroxylation. Our results indicate that edge functionalized ZBNNRs should present novel electronic and magnetic properties that may open the way for the design of new nano-spintronic devices.

**Acknowledgements**

This work was supported by the Israel Science Foundation (ISF) under grant No. 1313/08, the European Community's Seventh Framework Programme FP7/2007–2013 under grant agreement No. 249225, the Center for Nanoscience and Nanotechnology at Tel-Aviv University, and the Lise Meitner-Minerva Center for Computational Quantum Chemistry.

**References**

(1) Novoselov, K. S.; Geim, A. K.; Morozov, S. V.; Jiang, D.; Zhang, Y.; Dubonos, S. V.; Grigorieva, I. V.; Firsov, A. A. *Science* **2004**, *306*, 666.
(2) Berger, C.; Song, Z. M.; Li, X. B.; Wu, X. S.; Brown, N.; Naud, C.; Mayou, D.; Li, T. B.; Hass, J.; Marchenkov, A. N.; Conrad, E. H.; First, P. N.; de Heer, W. A. *Science* **2006**, *312*, 1191.
(3) Han, M. Y.; Ozyilmaz, B.; Zhang, Y. B.; Kim, P. *Phys. Rev. Lett.* **2007**, *98*, 206805.
(4) Cai, J. M.; Ruffieux, P.; Jaafar, R.; Bieri, M.; Braun, T.; Blankenburg, S.; Muoth, M.; Seitsonen, A. P.; Saleh, M.; Feng, X. L.; Mullen, K.; Fasel, R. *Nature* **2010**, *466*, 470.
(5) Hod, O.; Barone, V.; Peralta, J. E.; Scuseria, G. E. *Nano Lett.* **2007**, *7*, 2295.
(6) Stankovich, S.; Dikin, D. A.; Dommett, G. H. B.; Kohlhaas, K. M.; Zimney, E. J.; Stach, E. A.; Piner, R. D.; Nguyen, S. T.; Ruoff, R. S. *Nature* **2006**, *442*, 282.
(7) Schedin, F.; Geim, A. K.; Morozov, S. V.; Hill, E. W.; Blake, P.; Katsnelson, M. I.; Novoselov, K. S. *Nat. Mater.* **2007**, *6*, 652.
(8) Barbolina, I. I.; Novoselov, K. S.; Morozov, S. V.; Dubonos, S. V.; Missous, M.; Volkov, A. O.; Christian, D. A.; Grigorieva, I. V.; Geim, A. K. *Appl. Phys. Lett.* **2006**, *88*, 013901.
(9) Di, C. A.; Wei, D. C.; Yu, G.; Liu, Y. Q.; Guo, Y. L.; Zhu, D. B. *Adv. Mater.* **2008**, *20*, 3289.
(10) Kusakabe, K.; Maruyama, M. *Physical Review B* **2003**, *67*.
(11) Yamashiro, A.; Shimoi, Y.; Harigaya, K.; Wakabayashi, K. *Physical Review B* **2003**, *68*.
(12) Nakada, K.; Fujita, M.; Dresselhaus, G.; Dresselhaus, M. S. *Physical Review B* **1996**, *54*, 17954.
(13) Fujita, M.; Wakabayashi, K.; Nakada, K.; Kusakabe, K. *Journal of the Physical Society of Japan* **1996**, *65*, 1920.
(14) Son, Y. W.; Cohen, M. L.; Louie, S. G. *Nature* **2006**, *444*, 347.
(15) Gunlycke, D.; Li, J. W.; Mintmire, J. W.; White, C. T. *Nano Letters* **2010**, *10*, 3638.

(16) Ramprasad, R.; von Allmen, P.; Fonseca, L. R. C. *Physical Review B* **1999**, *60*, 6023.
(17) Cantele, G.; Lee, Y. S.; Ninno, D.; Marzari, N. *Nano Letters* **2009**, *9*, 3425.
(18) Gunlycke, D.; Li, J.; Mintmire, J. W.; White, C. T. *Applied Physics Letters* **2007**, *91*.
(19) Son, Y. W.; Cohen, M. L.; Louie, S. G. *Phys. Rev. Lett.* **2006**, *97*, 216803.
(20) Hod, O. *Journal of Chemical Theory and Computation* **2012**, *8*, 1360.
(21) Shi, Y. M.; Hamsen, C.; Jia, X. T.; Kim, K. K.; Reina, A.; Hofmann, M.; Hsu, A. L.; Zhang, K.; Li, H. N.; Juang, Z. Y.; Dresselhaus, M. S.; Li, L. J.; Kong, J. *Nano Letters* **2010**, *10*, 4134.
(22) Song, L.; Ci, L. J.; Lu, H.; Sorokin, P. B.; Jin, C. H.; Ni, J.; Kvashnin, A. G.; Kvashnin, D. G.; Lou, J.; Yakobson, B. I.; Ajayan, P. M. *Nano Letters* **2010**, *10*, 3209.
(23) Zeng, H. B.; Zhi, C. Y.; Zhang, Z. H.; Wei, X. L.; Wang, X. B.; Guo, W. L.; Bando, Y.; Golberg, D. *Nano Letters* **2010**, *10*, 5049.
(24) Gorbachev, R. V.; Riaz, I.; Nair, R. R.; Jalil, R.; Britnell, L.; Belle, B. D.; Hill, E. W.; Novoselov, K. S.; Watanabe, K.; Taniguchi, T.; Geim, A. K.; Blake, P. *Small* **2011**, *7*, 465.
(25) Tang, Q.; Zhou, Z.; Chen, Z. *The Journal of Physical Chemistry C* **2011**, *115*, 18531.
(26) Barone, V.; Peralta, J. E. *Nano Letters* **2008**, *8*, 2210.
(27) Park, C. H.; Louie, S. G. *Nano Letters* **2008**, *8*, 2200.
(28) Wang, Y. L.; Ding, Y.; Ni, J. *Physical Review B* **2010**, *81*.
(29) Lopez-Bezanilla, A.; Huang, J. S.; Terrones, H.; Sumpter, B. G. *Nano Letters* **2011**, *11*, 3267.
(30) Frisch, M. J. T., G. W.; Schlegel, H. B.; Scuseria, G. E.; Robb, M. A.; Cheeseman, J. R.;; Scalmani, G. B., V.; Mennucci, B.; Petersson, G. A.; Nakatsuji, H.; Caricato, M.; Li, X.;; Hratchian, H. P. I., A. F.; Bloino, J.; Zheng, G.; Sonnenberg, J. L.; Hada, M.; Ehara, M.;; Toyota, K. F., R.; Hasegawa, J.; Ishida, M.; Nakajima, T.; Honda, Y.; Kitao, O.; Nakai, H.;; Vreven, T. M., J. A.; Peralta, J. E.; Ogliaro, F.; Bearpark, M.; Heyd, J. J.; Brothers, E.;; Kudin, K. N. S., V. N.; Kobayashi, R.; Normand, J.; Raghavachari, K.; Rendell, A.; Burant, J.; C.; Iyengar, S. S. T., J.; Cossi, M.; Rega, N.; Millam, J. M.; Klene, M.; Knox, J. E.; Cross, J. B.;; Bakken, V. A., C.; Jaramillo, J.; Gomperts, R.; Stratmann, R. E.; Yazyev, O.; Austin, A. J.;; Cammi, R. P., C.; Ochterski, J. W.; Martin, R. L.; Morokuma, K.; Zakrzewski, V. G.; Voth, G.; A.; Salvador, P. D., J. J.; Dapprich, S.; Daniels, A. D.; Farkas; Foresman, J. B.; Ortiz, J. V.;; Cioslowski, J. F., D. J.  2009.
(31) Perdew, J. P.; Burke, K.; Ernzerhof, M. *Phys. Rev. Lett.* **1996**, *77*, 3865.
(32) Perdew, J. P.; Burke, K.; Ernzerhof, M. *Phys. Rev. Lett.* **1997**, *78*, 1396.
(33) Heyd, J.; Scuseria, G. E.; Ernzerhof, M. *J. Chem. Phys.* **2003**, *118*, 8207.
(34) Heyd, J.; Scuseria, G. E.; Ernzerhof, M. *J. Chem. Phys.* **2006**, *124*, 219906.
(35) Heyd, J.; Scuseria, G. E. *J. Chem. Phys.* **2004**, *120*, 7274.
(36) Heyd, J.; Scuseria, G. E. *J. Chem. Phys.* **2004**, *121*, 1187.
(37) Barone, V.; Hod, O.; Scuseria, G. E. *Nano Lett.* **2006**, *6*, 2748.
(38) Hod, O.; Peralta, J. E.; Scuseria, G. E. *Phys. Rev. B* **2007**, *76*, 233401.
(39) Hod, O.; Scuseria, G. E. *Acs Nano* **2008**, *2*, 2243.
(40) Barone, V.; Hod, O.; Peralta, J. E.; Scuseria, G. E. *Accounts Chem. Res.* **2011**, *44*, 269.
(41) Barone, V.; Hod, O.; Peralta, J. E.; Leszczynski, J., Ed.; Springer Netherlands: 2012, p 901.

(42)   This basis set consists of (3s2p1d) contracted Gaussian functions for C and Li and (2s 1p) for H

(43)   We note that none of the DFT exchange-correlation functional approximations that we consider provide a full description of the long-range correlation effects required for the appropriate description of hydrogen bonding. Therefore, we expect that a proper inclusion of such effects will further increase the calculated relative stability of the OH(II) oxidation scheme.

(44)   Dumitrica, T.; Hua, M.; Yakobson, B. I. *Physical Review B* **2004**, *70*.

(45)   We use a 68 k-point uniform grid for all calculations of the one-dimensional ZBNNRs studied.